\documentclass[%
 reprint,
 amsmath,amssymb,
 aps,
pra,
]{revtex4-2}

\usepackage{graphicx}% Include figure files
\usepackage{caption}
\usepackage{subcaption}
\usepackage{dcolumn}% Align table columns on decimal point
\usepackage{bm}% bold math
\usepackage[colorlinks, urlcolor = {blue}]{hyperref}% add hypertext capabilities
\usepackage{color}
\usepackage{ulem}

\begin{document}

%\preprint{APS/123-QED}

\title{Neutron Interferometers from Stacked Holographic Photopolymer Gratings}% Force line breaks with \\

 %\altaffiliation[Also at ]{}%Lines break automatically or can be forced with \\
\author{Saba Shams-Lahijani$^{1}$}% 
\author{Tobias Jenke$^2$}
\author{Jürgen Klepp$^1$} 
\email{juergen.klepp@univie.ac.at}
\author{Christian Pruner$^3$}
\author{Martin Fally$^1$}  
\email{martin.fally@univie.ac.at}
\affiliation{$^1$Faculty of Physics, University of Vienna, Boltzmanngasse 5, 1090 Vienna, Austria\\
$^2$Institut Laue-Langevin, 71 avenue des Martyrs, 38000 Grenoble, France\\ 
$^3$Department of Chemistry and Physics of Materials, University of Salzburg, Jakob-Haringer-Strasse 2a, 5020 Salzburg, Austria
}


%  \homepage{http://www.Second.institution.edu/~Charlie.Author}
% \affiliation{
%  Second institution and/or address\\
%  This line break forced% with \\
% }%
% \affiliation{
%  Third institution, the second for Charlie Author
% \affiliation{%
%  Authors' institution and/or address\\
%  This line break forced with \textbackslash\textbackslash
% }%

% \collaboration{CLEO Collaboration}%\noaffiliation

\date{\today}% It is always \today, today,

\begin{abstract} 
Long-wavelength neutron interferometry using discrete optical elements is notoriously challenging due to stringent alignment and stability requirements. Here, we introduce a monolithic double-Laue neutron interferometer fabricated from a stack of commercial \textit{Bayfol HX} photopolymer films. By recording holographic gratings simultaneously in multiple layers, we create a robust device that is inherently aligned, bypassing traditional stability problems. We demonstrate the device's function by observing the characteristic interference fringes in the diffracted intensity of both light and very cold neutrons. The interferometer is then used to perform in-situ characterization of the neutron beam's spectral profile, demonstrating its utility as a compact spectrometer. Our work establishes stacked holographic gratings as a simple, versatile, and powerful platform for matter-wave interferometry and metrology.

\end{abstract}

\maketitle

%\tableofcontents

\section{\label{sec:Intro}Introduction}

Diffractive optical elements (DOEs) are integral to modern technology, with widespread applications in telecommunications \cite{Buetuen-ama23}, laser optics \cite{Huang-oe13}, the automotive sector, and the rapidly growing field of augmented and virtual reality \cite{Guo-ole24}. A common fabrication method involves recording volume holographic patterns in photosensitive materials. This technology is mature, with many commercial products available, valued for their high sensitivity, thermal and mechanical stability, and ease of fabrication. Among these, the photopolymer \textit{Bayfol HX} stands out, primarily due to its exceptionally high saturated refractive-index modulation for light, $\Delta n_{1,L}\approx 3.4\times 10^{-2}$ \cite{Bruder-p17,Bruder-jpst09}. While in light optics DOEs are one of several established technologies for manipulating wave propagation alongside conventional refractive and reflective devices, the situation for thermal neutrons ($\lambda_\textrm{\tiny N}\approx 0.2$ nm) is fundamentally different.

In neutron optics, the refractive index $n_\textrm{\tiny N}$ is defined as
\begin{equation}\label{eq:nRI}
n_\textrm{\tiny N} \approx 1 - \frac{[b_c\rho]\lambda_\textrm{\tiny N}^2}{2\pi},
\end{equation}
where the material properties are described by the coherent scattering length density (SLD), $[b_c\rho]$ \cite{Sears-pb88}. The SLD is determined by the coherent scattering length $b_c$, which quantifies the isotope- and spin-specific interaction of an atom with neutrons, and the number density $\rho$. For all materials, typical values yield a refractive index very close to unity (e.g., $n_\textrm{\tiny N}-1\approx -10^{-6}$ at $\lambda_\textrm{\tiny N}=0.2$~nm). Consequently, significant refraction and reflection occur only at grazing incidence. This principle allows neutrons to be guided via total external reflection or with super-mirrors \cite{Mezei-cop76,Mezei-cop77}, but these methods typically require bulky and expensive equipment.

These constraints are somewhat relaxed for slow and cold neutron beams ($0.6 < \lambda_\textrm{\tiny N} < 10$ nm). For instance, double-reflection multilayer monochromators can reach up to $7^\circ$ for cold neutrons \cite{Hoeghoej-96}. As an alternative, particularly for manipulating very cold neutrons (VCN), neutron diffractive optical elements (nDOEs) in a transmission geometry offer a compelling solution due to their compact size and simpler handling. The past two decades of research into optimizing materials for nDOEs have led to significant achievements, including the successful operation of a cold-neutron interferometer \cite{Schellhorn-pb97,Pruner-nima06} and the fabrication of 50:50 beam splitters \cite{Fally-prl10}, three-port beam splitters \cite{Klepp-apl12a}, and mirrors \cite{Klepp-apl12}. Current research focuses on achieving high diffraction efficiency for nDOEs that can operate with poorly collimated, broadband beams. This capability is crucial for utilizing the full spectrum of typical VCN sources and is essential for the practical operation of VCN interferometers \cite{Tomita-pra20,Hadden-spie22}.

In this paper, we introduce a novel approach for fabricating nDOEs: stacked holographic gratings (multilayer gratings) recorded in \textit{Bayfol HX} photopolymer. This concept is inspired by similar techniques developed for light optics applications \cite{Pen-qe10,Bruder-spie20}. We present an experimental investigation of these stacked gratings, characterizing their diffraction properties for both light and neutrons.

\section{Materials and Methods}

\begin{figure*}
\centering
% \frame{
\includegraphics[height=4.5cm]{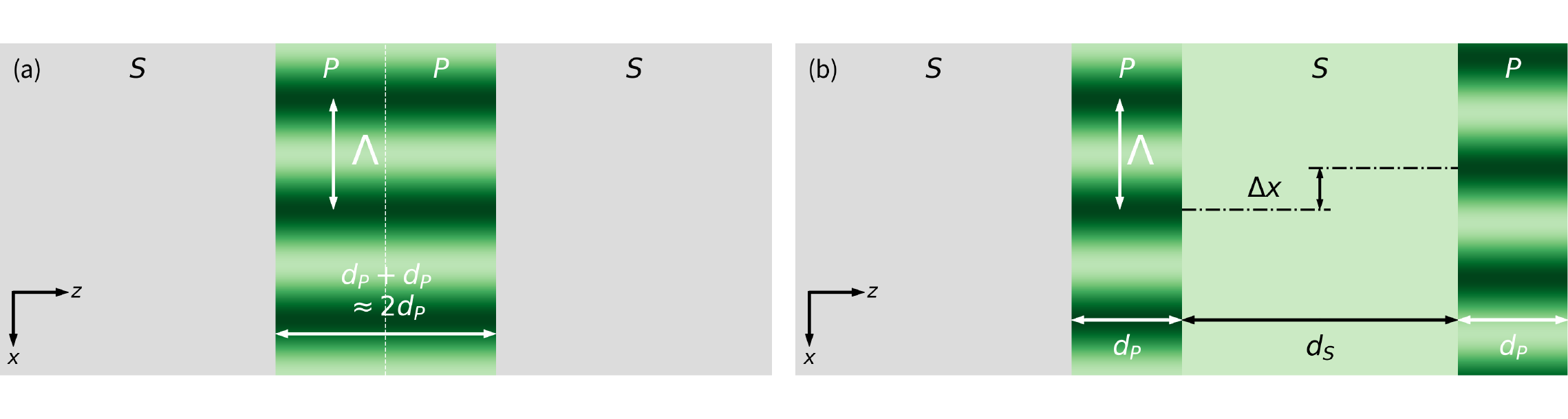}
%   }
\caption{\label{fig:geo} Stack of gratings. (a) Configuration \textsf{SPPS}, (b) configuration \textsf{SPSP}. For P layers, the recorded grating profiles are schematically shown.}
\end{figure*}

\subsection{Samples and holographic recording} 

The grating stacks were prepared using commercial \textit{Bayfol HX} photopolymer films. Each film is composed of three distinct layers: a protective cover layer, which is removed prior to recording; the photosensitive polymer layer with a nominal thickness of $d_P\approx 16~\mu$m; and a transparent substrate layer with a thickness of $d_S\approx 50~\mu$m.

To create two distinct sample types, we stacked two films after removing their protective cover layers. This process resulted in the following four-layer configurations:
\begin{enumerate}
 \item[\textit{(i)}] substrate-polymer-polymer-substrate (\textsf{SPPS}), and
 \item[\textit{(ii)}] substrate-polymer-substrate-polymer (\textsf{SPSP}).
\end{enumerate}

In configuration \textsf{SPPS}, the two adjacent polymer layers form a \textbf{single, thick grating} with an effective thickness of $2d_P\approx 32~\mu$m upon holographic recording. This grating is sandwiched between the two substrate layers, which are optically inert and remain unmodified by the recording light. Conversely, configuration \textsf{SPSP} results in a double-grating structure. Here, two individual gratings, each with a thickness of approximately $d_P$, are formed in the separate polymer layers. These gratings are separated by the central, optically inert substrate layer of thickness $d_S$.
The resulting grating structures are schematically depicted in Fig.\,\ref{fig:geo}. While the outer substrate layers in configuration \textsf{SPPS} enhance the sample's mechanical stability, they do not contribute to its function as a DOE.

This work focuses on the \textsf{SPSP} configuration, which effectively forms a double-Laue (LL) interferometer. This structure is analogous to established interferometers for X-rays \cite{Bonse-apl65,Fukamachi-jjap04} and neutrons \cite{Arthur-prb85,Heacock-aca19}. While this study is limited to a double-layer stack, more complex devices can be constructed by stacking additional photopolymer and substrate layers, as explored via simulation in Ref.~\cite{Lahijani-spie23}. Further details on the fabrication of (n)DOEs can also be found in that work.

Holographic recording was performed using a single-longitudinal-mode laser (Coherent Genesis CX-514 SLM) operating at a wavelength of $\lambda_\text{rec}=514~$nm. Transmission gratings were recorded simultaneously within both photopolymer (P) layers of the stack by interfering two $s$-polarized plane waves. The beams enclosed an angle of $62.8^\circ$, which created a sinusoidal interference pattern with a period of $\Lambda\approx 493~$nm.

In response to light exposure, \textit{Bayfol HX} undergoes a change in its refractive index. This change can include higher-order harmonics \cite{Bruder-spie15}, leading to a spatially modulated refractive index for light, $n(x)$, which can be described by the Fourier series:
\begin{equation}\label{eq:ref_in}
n(x)=n_0+\sum_{m\geq 0} \Delta n_m\cos\left(m\frac{2\pi}{\Lambda}x+\varphi_m\right),
\end{equation}
where $n_0$ is the bulk refractive index of the unexposed material. The terms $\Delta n_m$ and $\varphi_m$ represent the amplitude and phase of the $m$-th order harmonic of the refractive index modulation, respectively. Slight variations in the recording process or material properties (e.g., thickness, sensitivity) can result in minor differences between the two gratings, such as deviations in their thickness, modulation amplitude, or their relative lateral alignment.

The refractive index for neutrons, $n_\text{\tiny N}(x)$, is derived from the corresponding spatial modulation of the SLD via Eq.\,(\ref{eq:nRI}):
\begin{equation}
 n_\text{\tiny N}(x)\approx 1-\frac{\lambda_\text{\tiny N}^2}{2\pi}b_c\left[\rho_0+
 \sum_{m\geq 0} \Delta \rho_m\cos\left(m\frac{2\pi}{\Lambda}x+\varphi_m\right)\right].\label{eq:n_neutron}
\end{equation}
For \textit{Bayfol HX} gratings, the contribution of higher-order harmonics to the neutron interaction is negligible. Therefore, Eq.\,(\ref{eq:n_neutron}) simplifies to a purely sinusoidal modulation:
\begin{equation}
	n_\text{\tiny N}(x)\approx 1-\frac{\lambda_\text{\tiny N}^2}{2\pi}b_c\left[\rho_0+\Delta\rho_1\cos\left(\frac{2\pi}{\Lambda}x\right)\right].\label{eq:n_neutron1}
\end{equation}

\subsection{The \textsf{SPSP} configuration as an LL-interferometer}

\begin{figure*}
\frame{\includegraphics[width=12cm,trim=0 140 0 140,clip]{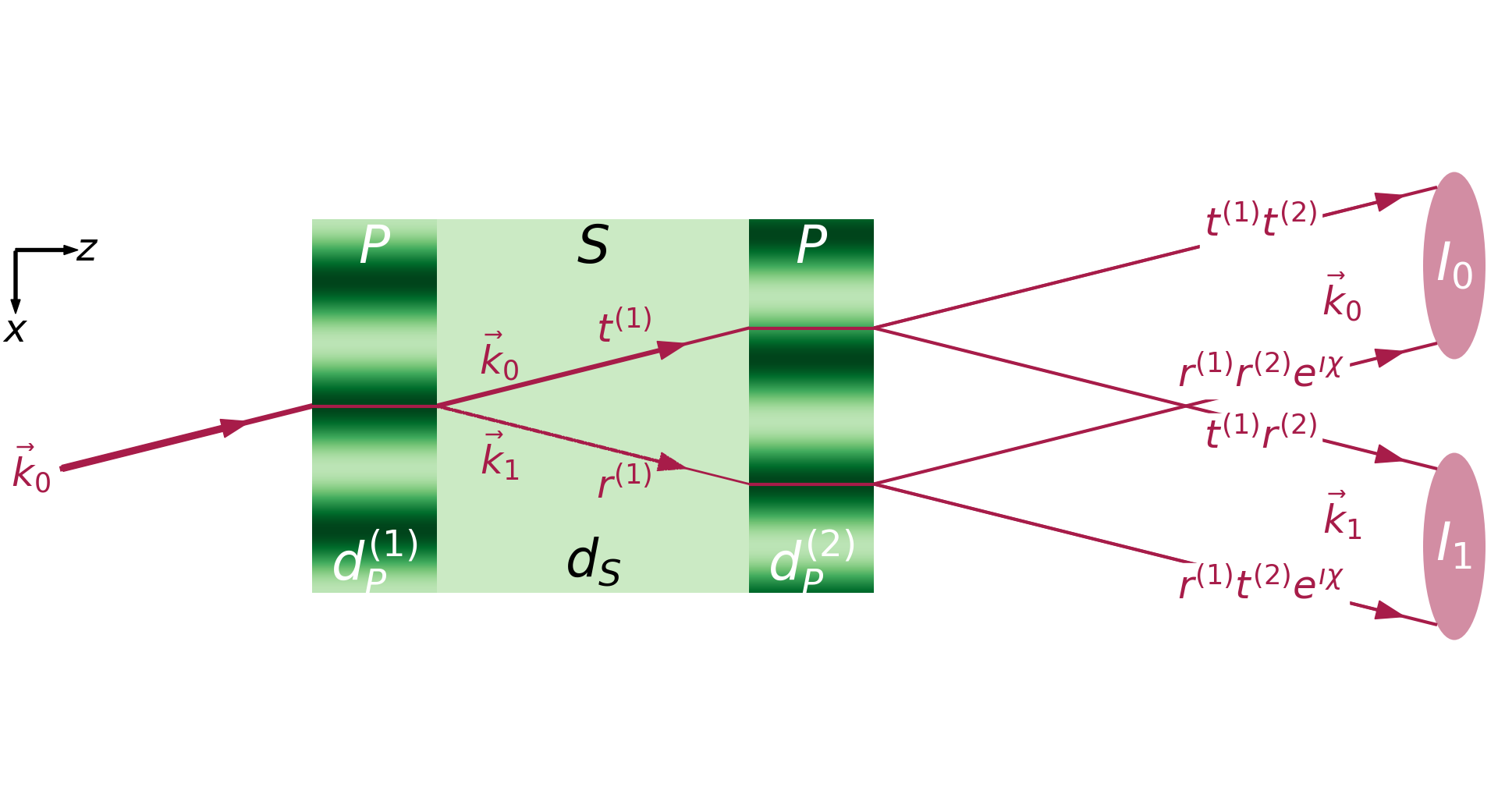}}
\caption{\label{fig:ifsketch} Schematic of the LL-interferometer (not to scale). An incident in-Bragg beam, which is approximately a plane wave with wavevector $\vec k_0$, impinges on the first grating. This will result in two waves with amplitudes $t^\text{\scriptsize{(1)}}, r^\text{\scriptsize{(1)}}$ and wavevectors $\vec k_0, \vec k_1$. Both pass the substrate layer and thereby accumulate additional phases, their difference being $\chi$. Finally each of the waves is diffracted from the second grating leading to multiplication of amplitudes by factors $t^\text{\scriptsize{(2)}}, r^\text{\scriptsize{(2)}}$. The output intensities, $I_0$ and $I_1$, arise from the coherent superpositions of two distinct beam paths: Given the small Bragg angle and thin substrate layer $d_S$, these paths overlap spatially, resulting in a two-beam interference pattern in each diffraction channel.  
}	
\end{figure*}

The \textsf{SPSP} configuration forms an LL-interferometer with coherent (interfering) beams along the directions $\vec k_0$ of zero order  and $\vec k_{1}$ of first order  diffraction, respectively. The geometry of this setup is depicted schematically in Fig.\,\ref{fig:ifsketch}. 

The complex-valued amplitudes for transmission and diffraction at the first and second gratings are denoted by $t^\text{\scriptsize{(1)}},r^\text{\scriptsize{(1)}}$ and $t^\text{\scriptsize{(2)}}, r^\text{\scriptsize{(2)}}$, respectively. These amplitudes depend on both the grating strength and the angle of incidence. In the far field, the intensity along the zeroth-order direction ($\vec k_0$) is given by 
\begin{equation*}
I_0\propto|t^\text{\scriptsize{(1)}}t^\text{\scriptsize{(2)}}+r^\text{\scriptsize{(1)}}r^\text{\scriptsize{(2)}}e^{\imath\chi}|^2, 
\end{equation*} 
while the intensity along the first-order direction ($\vec k_1$) is 
\begin{equation*}
I_1\propto|t^\text{\scriptsize{(1)}}r^\text{\scriptsize{(2)}}+r^\text{\scriptsize{(1)}}t^\text{\scriptsize{(2)}}e^{\imath\chi}|^2. 
\end{equation*}

The phase shift $\chi$ arises from the optical path length difference between the two beam paths within the intermediate substrate. This phase can be controlled, for example, by rotating the entire interferometer about an axis perpendicular to the $x-z$ plane. It is noteworthy that large-scale LL-interferometers for neutrons, which utilize thick gratings and narrow incident beams, were demonstrated several decades ago \cite{Kikuta-jpsj78}.

\subsection{Light Optical Diffraction (readout)} 

The angular dependence of the diffraction efficiency, $\eta_{\pm 1}(\Theta)$, was characterized for the \textsf{SPSP} sample configuration. For these measurements, we used a 3\,mm wide, $s$-polarized He-Ne laser beam operating at a wavelength of $\lambda=543.5$\,nm.
The diffraction efficiency is defined as
$
\eta_{\pm 1}(\Theta) = I_{\pm 1}(\Theta) / \sum_j I_j(\Theta),
$
where $\Theta$ is the angle of incidence. The term $I_j$ represents the optical power of the $j$-th diffraction order. The power of each diffracted beam was measured using photodiodes.

\subsection{Neutron Optical Diffraction (readout)} 

Neutron diffraction measurements were conducted at the VCN beamline PF2/VCN at the Institut Laue-Langevin (ILL) in Grenoble, France. The beam is characterized by a mean wavelength of $\overline{\lambda}_\text{\tiny N}\approx 5.5$\,nm and a broad spectral distribution, which can be described by an exponentially modified Gaussian function with an extended tail towards longer wavelengths \cite{Tomita-pra20,Blaickner-nima19}.

To achieve a beam collimation of approximately 2\,mrad, we used a setup of two slits (3\,mm and 1\,mm wide, respectively) separated by a distance of 188\,cm upstream of the sample. The detector was placed 140\,cm downstream from the sample. This distance ensured that the zeroth- and first-order diffracted beams, separated by an angle of $2\Theta_{\pm 1}\approx \lambda_\text{\tiny N}/\Lambda\approx 11$~mrad, could be clearly resolved. On the 2D detector, which comprises a $128\times 128$ pixel array with a pixel size of $2\times 2\,\textrm{mm}^2$, this corresponded to a spatial separation of approximately 7.7 pixels. To minimize scattering losses from air, the entire neutron flight path from the entrance slit to the detector was enclosed in tubes filled with He gas.
% % % %
% % Wavelength checked using "mirror": slow tail 7.9 nm, fast 5.0 nm, average: 5.45 nm
% % % % 

For data analysis, the detector counts were integrated within predefined regions of interest (ROIs) corresponding to the zeroth- and first-order diffraction peaks for each measured angle of incidence. This process allowed us to determine the angular dependence of the diffraction efficiency, $\eta_{\pm 1}(\Theta)$, which serves to characterize the neutron-optical properties of the nDOEs.

\section{Modelling the Angular Response of a Grating Stack\label{sec:Model}}

For modeling light diffraction, a two-coupled-wave (2CW) approach is sufficient. This is because, under Bragg-matching conditions, only the zeroth-order and one first-order beam propagate significantly through the grating. For a single grating, we therefore apply the established solutions to the wave equation in periodic media, developed in various forms by authors such as Kogelnik \cite{Kogelnik-atj69} and Uchida \cite{Uchida-josa73}. Our model is specifically based on the beta-value method. This approach was extended to multilayer grating systems in Ref.~\cite{Au-jmo87} using a transfer-matrix formalism \cite{Born-19}. More recently, this model was adapted in Ref.~\cite{Lahijani-spie23} to include a lateral shift $\Delta x$ between the gratings in configuration \textsf{SPSP}, albeit under the simplifying assumption of two identical, lossless gratings (see Eqns.\,(2)-(5) therein). Such a shift can arise from non-ideal recording conditions, for instance, a minor misalignment relative to the symmetric geometry shown in Fig.\,\ref{fig:geo}. In this work, we lift the restriction of identical gratings and present a generalized form of these equations in Appendix~\ref{sec:appendix}.

\newcommand{\bcr}{b_c\Delta\rho_{1}}

For modeling neutron diffraction, however, the situation is fundamentally different. Because the grating period is much larger than the neutron wavelength ($\Lambda\gg\lambda_\text{\tiny N}$), multiple diffraction orders propagate simultaneously. This requires extending the model to multi-wave coupling \cite{Tomita-pra20,Klepp-jpcs16}. Specifically, we employ a first-order theory that includes three coupled waves (3CW). For simplicity, we assume that the SLD modulation, $\bcr$, is identical for both grating layers. We solve the system of three linear differential equations for each layer and use a transfer-matrix approach, where the complex output amplitudes from layer $\ell$ serve as the input for layer $\ell+1$. The final amplitudes ($S_{-1},S_0,S_{+1}$) of the three propagating waves are thus given by the product of the layer matrices:
\begin{equation}
 \left[
 \begin{array}{c}
  S_{-1}\\
  S_{0}\\
  S_{+1}
 \end{array}
\right]_{3}
=\prod\limits_{\ell=1}\limits^3 M_\ell
\left[
 \begin{array}{c}
  0\\
  1\\
  0
 \end{array}
\right].
\end{equation}
Here, the matrix $M_\ell\in\mathbb{C}^{(3,3)}$ represents the complex amplitude transfer function for layer $\ell$. This model neglects absorption and reflection, an approximation that is well-justified for neutrons due to their weak interaction with matter.

To validate this 3CW approximation, we compared its results to those from a rigorous coupled-wave analysis (RCWA) that includes second derivatives and higher diffraction orders, following the methods of Refs.\,\cite{Moharam-josaa95,Moharam-josaa95a,Neipp-oc04}. The results from both models were identical within the numerical precision of the calculations. This excellent agreement is expected because several conditions for the validity of the simpler model are met: the bulk refractive index for neutrons is nearly unity, which minimizes reflection at interfaces; the diffraction angles are small; the gratings are sinusoidal (i.e., higher-order Fourier coefficients are negligible); and the gratings are unslanted.

\section{Experimental Results}
\subsection{Single Grating (\textsf{SPPS})}

Previous characterization of the \textsf{SPPS} sample with light at $\lambda=632.8$~nm revealed a thickness-averaged refractive-index modulation of $\Delta n_1 \approx 1.6\times 10^{-2}$ and a total grating thickness of $d\approx 31~\mu$m. The modulation exhibited a moderate exponential decay along the grating depth ($1/e\text{-length}\approx 25~\mu$m) \cite{Lahijani-spie23}. 

\begin{figure*}
\centering
\includegraphics[width=12cm]{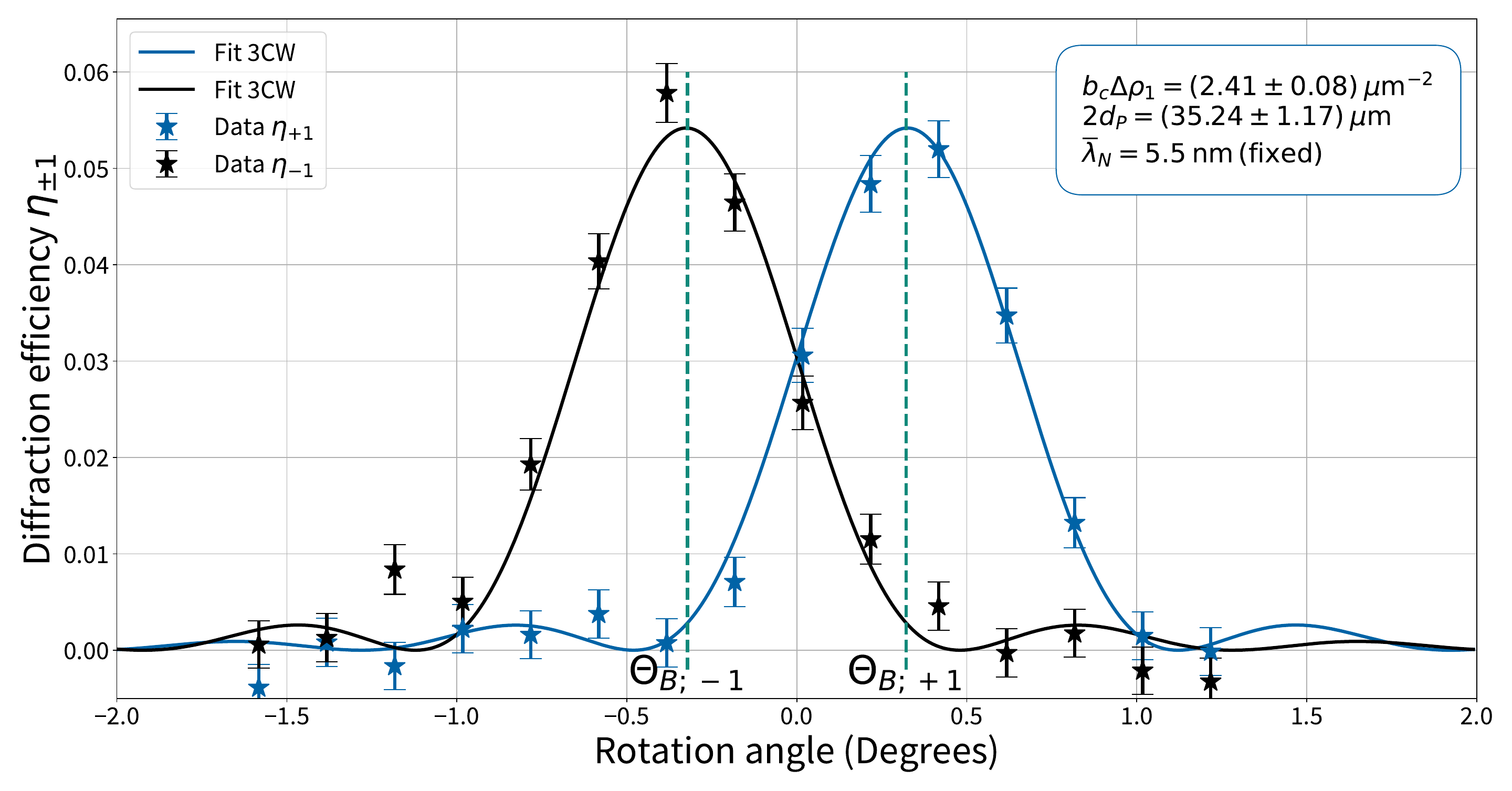}
\caption{Angular dependence of the diffraction efficiency for the \textsf{SPPS} stack, measured with a VCN beam ($\overline{\lambda}_\text{\tiny N}=5.5$~nm). The data are available from the ILL data portal \cite{Klepp-ILI20}. To increase the effective thickness, the grating was tilted by $\zeta\approx 35^\circ$ about an axis parallel to the grating vector, such that the effective thickness becomes $d_\text{eff}=2d_P/\cos\zeta$. The solid lines are fits based on a first-order 3-wave coupling theory \cite{Klepp-jpcs16}. \label{fig:n_SGS}
}
\end{figure*}

While in-plane neutron diffraction for this sample was reported in \cite{Lahijani-spie23}, here we aimed to increase the diffraction efficiency by tilting the sample. The results, obtained at a rotational tilt of $\zeta=35^\circ$ about an axis parallel to the grating vector, are shown in Fig.\,\ref{fig:n_SGS}. The data are well-described by the first-order 3-wave coupling model \cite{Klepp-jpcs16}.

Due to the grating's small thickness ($d_P$) and the resulting low angular selectivity (FWHM $\approx \Lambda/d_P\approx 0.7^\circ$), the effect of the broad VCN wavelength distribution was negligible in the analysis. This significantly simplifies the data evaluation, as it avoids the computationally expensive step of averaging the theoretical model over the beam's spectral distribution.

%%%%%%%%%%%%%%%%%%%%%%%%%%%%%%%%%%%%%%%%%

\subsection{Double-Grating (\textsf{SPSP}) LL-Interferometer} 

\begin{figure*}
\centering
\includegraphics[width=14cm,trim=0 0 0 0,clip]{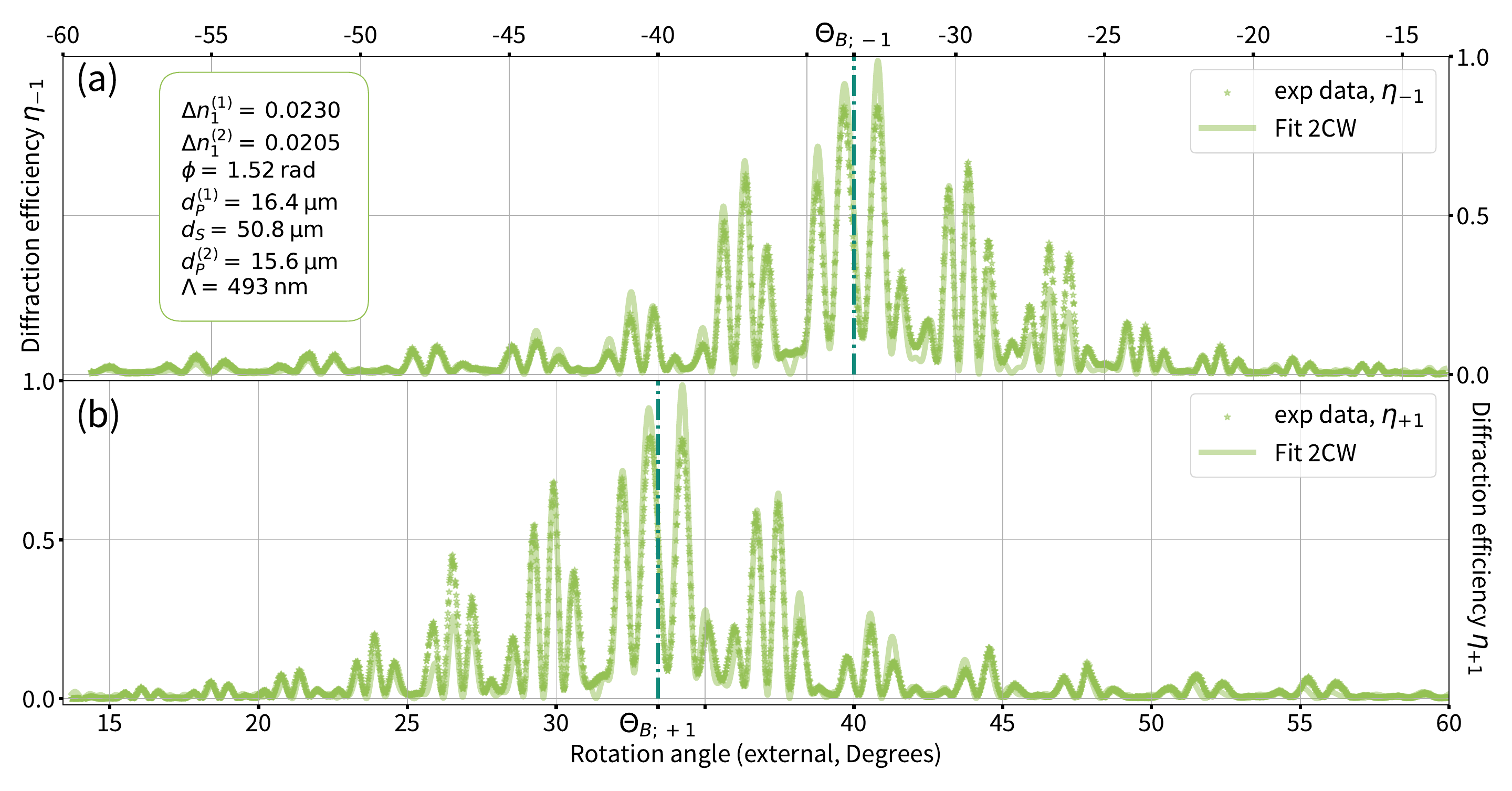}
\caption{
Angular dependence of the diffraction efficiency for the \textsf{SPSP} configuration, measured with light at a wavelength of $\lambda=543.5$\,nm. Upper panel: (\textbf{a}) minus-first order, $\eta_{-1}(\Theta)$. Lower panel: (\textbf{b}) plus-first order, $\eta_{+1}(\Theta)$. The solid lines represent fits based on a 3-layer, two-coupled-wave (2CW) model using the function given in Eq.\,(\ref{eq:2CW-3L}).
\label{fig:543_3GSG}
}
\end{figure*}

Figure~\ref{fig:543_3GSG} shows the measured angular dependence of the $\pm 1$ order diffraction efficiencies for the \textsf{SPSP} LL-interferometer, read out at a wavelength of $\lambda=543.5$\,nm.

\begin{figure*}
	\centering
	\includegraphics[width=14cm,bb=0 0 1382 723]{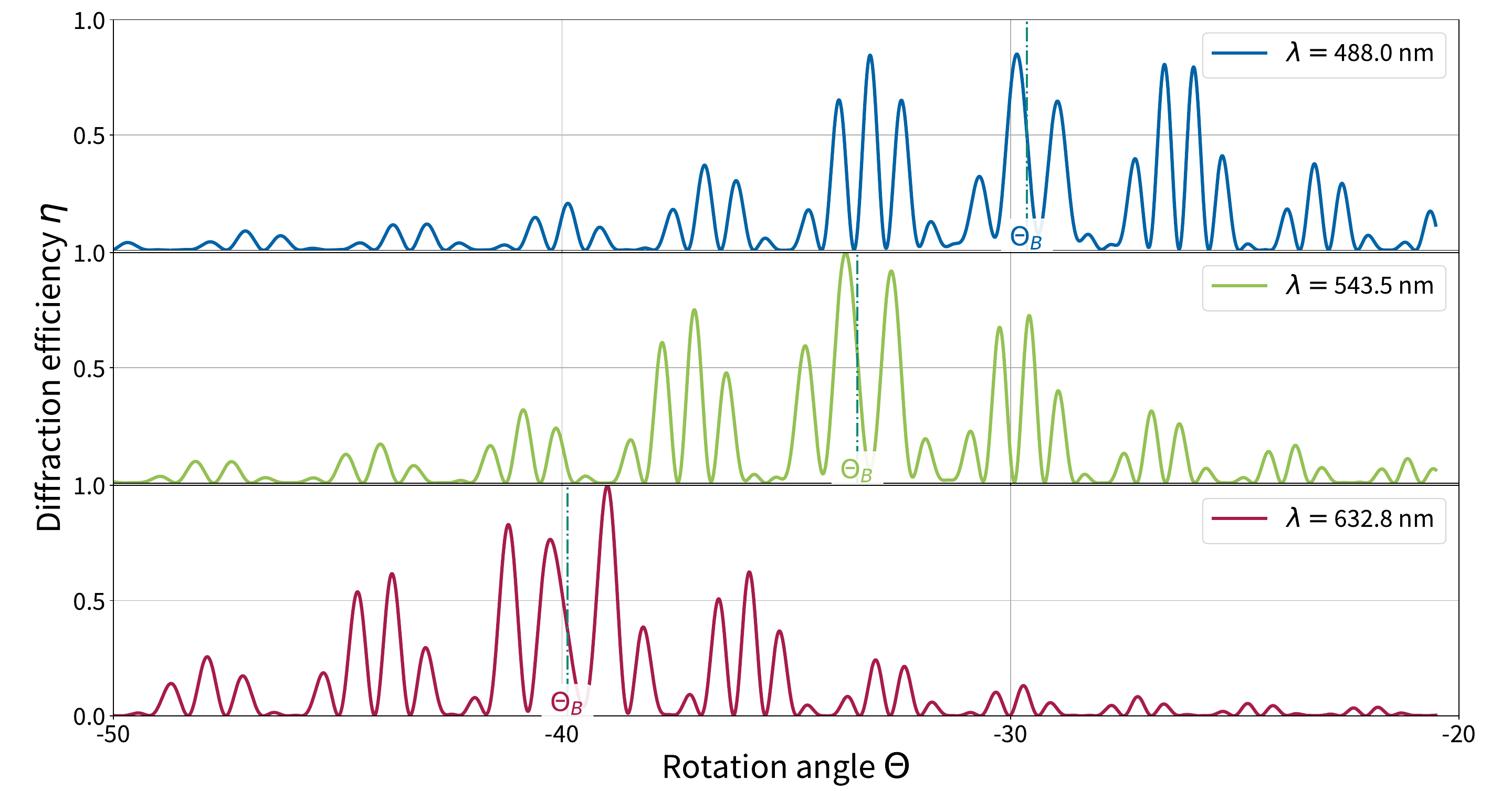}
% WL456.pdf: 1382x723 px, 72dpi, 48.75x25.51 cm, bb=0 0 1382 723
\caption{\label{fig:456} Simulation of the angular dependence of the diffraction efficiency for the \textsf{SPSP} configuration at three different wavelengths. The parameters used are those obtained from fitting the data at $\lambda=543.5$\,nm (see Fig.\,\ref{fig:543_3GSG}).}
\end{figure*}

As the beams traverse the intermediate \textsf{S}-layer, the path length difference $\Delta$ between the forward-diffracted path and the diffracted-transmitted path (see Fig.\,\ref{fig:ifsketch}) induces a relative phase shift $\chi$. This phase shift modulates the measured intensity. In our experiment, $\chi(\Theta)=\Delta(\Theta)2\pi n_S/\lambda$ is varied by rotating the entire interferometer, thereby detuning the angle of incidence $\Theta$ from the Bragg condition. Alternatively, $\chi$ could be modified by varying the wavelength. This wavelength-dependent behavior is evident when comparing the data in Fig.\,\ref{fig:543_3GSG} ($\lambda=543.5\,\text{nm}$) with those presented in Fig.\,3 of Ref.\,\cite{Lahijani-spie23} ($\lambda=632.8\,\text{nm}$). Figure~\ref{fig:456} shows a simulation for three different wavelengths using the parameters obtained from our fits.

The refractive index of the \textsf{S} layer is known a priori \cite{BayfolX-16}, as is the laser wavelength. Fitting the data with the function from Eq.\,\ref{eq:2CW-3L} accurately models the non-zero minima observed in the diffraction efficiency near the Bragg angle. Our model accounts for potential asymmetries, such as differing grating thicknesses $d_P^{\text{\scriptsize{(1)}},\text{\scriptsize{(2)}}}$ and refractive-index modulations $\Delta n_1^{\text{\scriptsize{(1)}},\text{\scriptsize{(2)}}}$, as well as a relative lateral shift $\phi$ between the gratings. Such imperfections can arise from non-ideal recording conditions, like a slightly slanted recording geometry. It is important to note that these features cannot be reproduced by the simplified model of Eq.~(2) in Ref.\,\cite{Lahijani-spie23}. This comprehensive model allows us to extract not only the grating parameters but also the thickness of the substrate layer, $d_S$, and the grating period, $\Lambda$. 

The light optical diffraction data reveal intensity oscillations composed of both fast and slow components. These oscillations are characteristic of the interference between waves diffracted by the two stacked gratings, and their period depends on the separation distance $d_S$ and the refractive index $n_S$. The fast oscillation period decreases with decreasing mutual distance $d_S$. Resolving similar oscillations in the neutron diffraction data imposes stringent experimental requirements: (1) a sufficiently small angular step size, (2) high beam collimation (on the order of a few mrad), and (3) a sufficiently narrow wavelength distribution ($\Delta\lambda/\lambda<0.3$).

% \textcolor{blue}{Frage woher das $\Delta x$ kommt:} Ehrliche Antwort ist - ich weiss es nicht; z.B. von einer minimalen Abweichung von der Bragg-Bedingung beim Schreiben, von minimalen Dickenunterschieden der Substrat-Folien oder deren Brechwerten\ldots\textcolor{red}{Ende Frage.}

Accordingly, we set the experimental parameters as follows: an angular step size of $\delta\Theta=0.1^\circ$ and a beam divergence of approximately 2\,mrad. The neutron counts were integrated for 20 minutes at each angular position. Unlike the \textsf{SPPS} case, the data analysis for the \textsf{SPSP} configuration requires accounting for the broad wavelength distribution. This is achieved by weighting the calculated diffraction efficiencies with an exponentially modified Gaussian (EMG) distribution, which is a convolution of an exponential and a normal distribution \cite{Grushka-ac72}:
\begin{eqnarray*}
\text{EMG}(\lambda;\mu,\sigma,\gamma)&=&\frac{\gamma}{2}\exp \left\{{\frac {\gamma }{2}}(2\mu +\gamma \sigma ^{2}-2\lambda)\right\}\\
&&\times\left[1-\text{erf}\left(\frac{\mu+\gamma\sigma^2-\lambda}{\sqrt{2}\sigma}
\right)\right]
% \\
% 1-\text{erf}(x)&=&\frac{2}{\sqrt{\pi}}\int\limits_x^{\infty}e^{-t^2}dt\nonumber
\end{eqnarray*}
with $\gamma, \mu, \sigma$ as free parameters. This parametrization is an alternative to that used in Ref.\,\cite{Abele-nima06} and, like it, does not imply a specific physical model for the beam spectrum. 

\begin{figure*}
\centering
\includegraphics[width=12cm]{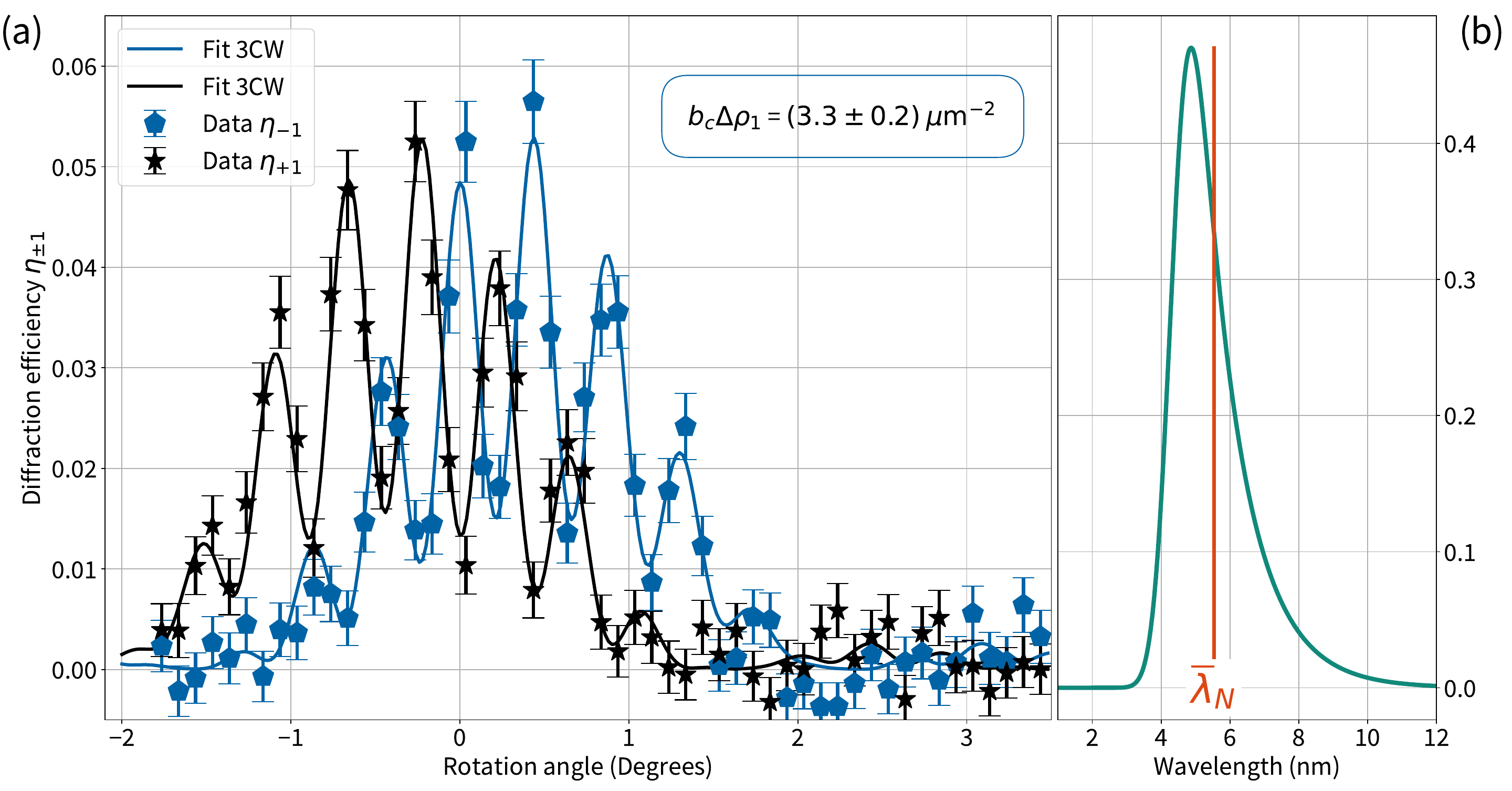}
% %%%%%%%%%%%%%%%%%%%%%%%%%/home/fallym4/MyPapers/24/BayFol_Light_Neutron/python_/3CW_PG_Multilayer_2024_002.py
\caption{\textbf{(a)}: Angular dependence of the $\pm$ first diffraction efficiencies for the \textsf{SPSP} configuration using VCNs. The lines represent fits to a first-order, three-wave coupling model that includes the VCN beam's wavelength distribution. Data are available from the ILL data portal \cite{Klepp-ILI20}.
%%% mean_lambda= (5.54 +/- 0.29) nm
\textbf{(b)} The shape of the EMG distribution as obtained from the fit to our data.
% \\\textcolor{red}{Frei fitten?; alternativ die Werte von Fig. 4?}
\label{fig:n_GSG}}
\end{figure*}

Figure~\ref{fig:n_GSG}(a) shows the angular dependence of the $\pm 1$ order neutron diffraction efficiencies. While the peak efficiencies of less than 0.1 are modest, as expected from the \textsf{SPPS} results, the data clearly exhibit the characteristic rapid oscillatory behavior of the LL-interferometer. To extract the neutron-optical parameters, the data were fitted using the 3CW model detailed in Section\,\ref{sec:Model}. The resulting fits are shown as solid lines in Figure~\ref{fig:n_GSG}(a), with the best-fit parameters listed in the inset. The corresponding VCN wavelength distribution, derived from the fit, is shown in Figure\,\ref{fig:n_GSG}(b) and is consistent with the known beamline spectrum published in Ref.\,\cite{Oda-nima17}. 

The fitted SLD modulation, $\bcr=3.3\,\mu\text{m}^{-2}$, is approximately 30\,\% higher than that obtained for the \textsf{SPPS} geometry (see Fig.\,\ref{fig:n_SGS} of the present manuscript and Fig.\,2 of Ref.\,\cite{Lahijani-spie23}). We attribute this improvement to optimized recording conditions for this sample. Nevertheless, this value remains about three times lower than the benchmark values reported for state-of-the-art holographic materials in neutron optics \cite{Fally-prl06,Hadden-apl24,Hadden-sr25}. Consequently, the maximum achievable diffraction efficiency with \textit{Bayfol HX} is roughly an order of magnitude lower. The thicknesses obtained for the gratings and the substrate correspond well to the expected values.

\section{Discussion and Conclusion}

Neutron interferometry, a cornerstone of experimental quantum mechanics, often relies on perfect-crystal designs. Outside of monolithic designs, neutron interferometers for long wavelengths, built from individual optical elements, fall prey to notorious alignment and stability challenges, demanding precision that is difficult to achieve and maintain. In this work, we introduce an elegant and powerful solution: monolithic interferometers fabricated from stacked, commercially available photopolymer films.
Our method of simultaneous holographic recording in multiple layers circumvents these stability challenges entirely, creating a robust and inherently stable device from a single exposure. This approach transforms a notoriously difficult experimental technique into a straightforward fabrication process. A key benefit of this fabrication approach is its versatility. The substrate can be replaced or even removed \cite{Bruder-spie20}, enabling the fabrication of stacks where a particular spacer material itself might be the object of investigation. Furthermore, this method allows for the assembly of more complex multilayer stacks, such as compact triple-Laue interferometers, provided the substrates are transparent to light. 
Clearly, a critical drawback of our monolithic approach is its inherent incompatibility with experiments that require inserting a sample into the interferometer's beam path. This limitation precludes a broad class of experiments, including those that use specialized phase shifters, large samples, or applied electric and magnetic fields. Adapting the fabrication process to integrate such components poses a formidable—and in some cases, insurmountable—challenge.

We have shown that grating stacks can be fabricated from \textit{Bayfol HX} films, each comprising a photosensitive polymer layer (\textsf{P}) and an optically inert substrate layer (\textsf{S}). Depending on the stacking sequence—\textsf{SPPS} or \textsf{SPSP}—either a single grating of double thickness or a double-grating structure is formed. While the former configuration simply increases the effective thickness, the latter constitutes a double-Laue (LL) interferometer for both light and neutrons. 

First, we analyzed the light-optical diffraction in the Bragg regime. 
% , i.e. only two diffracted waves with considerable amplitudes propagate when Bragg's condition is approximately fulfilled 
As shown in Fig.\,\ref{fig:543_3GSG}, the data exhibit oscillations with both a fast and a slow component. The slow component is governed by the thickness $d_P$ and refractive-index modulation $\Delta n_1$ of the individual gratings, while the fast component is sensitive to the spacer thickness $d_S$ and its refractive index $n_S$. By fitting the data with our advanced model (see Appendix\,\ref{sec:appendix}), we extracted these parameters. The obtained thicknesses are in excellent agreement with the manufacturer's specifications \cite{BayfolX-16}: $d_P=(16\pm 2)\,\mu\text{m}$ and $d_S=(50\pm 2)\,\mu\text{m}$.  

As anticipated, probing the \textsf{SPSP} grating stack with the broadband VCN beam also revealed a fast oscillatory behavior in the diffracted intensity as a function of angle, while the corresponding slow oscillation was not observable within the limited angular range of the neutron measurement. The presence of these fast oscillations is a clear signature that the double-grating configuration functions as intended: the neutron beam is coherently split and subsequently recombined, producing interference fringes that depend on the relative phase accumulated between the beam paths (Fig.\,\ref{fig:ifsketch}).

Since, in our experiment, the exact wavelength distribution of the VCN beam was unknown a priori, we used the LL-interferometer itself as a characterization tool. Its spectral profile can be parameterized by an EMG function \cite{Blaickner-nima19}. By fitting the measured angular dependence of the diffraction efficiency with our 3CW model, we were able to retrieve the three characteristic parameters of the distribution. While the overall diffraction efficiency is modest, the key structural parameters extracted from the fit, such as the layer thicknesses and the relative grating shift, are consistent with those obtained from the light-optical characterization. The resulting fit is shown in Fig.\,\ref{fig:n_GSG}(a), and the deduced VCN beam distribution is presented in Fig.\,\ref{fig:n_GSG}(b). This demonstrates the device's capability to serve as a compact spectrometer for characterizing the neutron beam itself.

% A final remark concerning the unusual geometry of the LL-interferometer discussed here is required. LLL-interferometers make use of three gratings, in which the forward diffracted beams at the second grating are blocked or leave the interferometer as they are spatially separated from the diffracted beams. In case that they contribute to the signal, the are even termed as parasitic beams \cite{Rauch-15}. 

In conclusion, we have demonstrated that by holographically patterning an \textsf{SPSP} stack of commercial \textit{Bayfol HX} films, a functional LL-interferometer for both light and neutrons can be readily fabricated. The device's characteristic sensitivity to phase changes was shown via rocking curve measurements. This capability allowed us to retrieve the spectral shape of the VCN beam, a parameter that would otherwise be difficult to determine due to the beam's complex profile at the time of the experiment \cite{Oda-nima17}. Our findings establish stacked photopolymer gratings as a versatile and robust platform for the next generation of compact neutron optical devices. The ability to rapidly design and fabricate custom multilayer stacks on a per-experiment basis opens new avenues for fundamental research in quantum optics and metrology with neutron matter waves.

% Authors should discuss the results and how they can be interpreted from the persp^h directions may also be highlighted.

% \begin{itemize}
%  \item Light, wavelength, 3L vs 1L
%  \item Ageing
%  \item Neutron, wavelength dist, 3L=>perspective
% \end{itemize}

%%%%%%%%%%%%%%%%%%%%%%%%%%%%%%%%%%%%%%%%%%
%%%%%%%%%%%%%%%%%%%%%%%%%%%%%%%%%%%%%%%%%%
% \authorcontributions{Authors contributions are as follows: ``Conceptualization, J.K. and M.F.; sample and grating preparation, S.S.-L., light optical experiments, S.S.-L. and M.F.; neutron optical experiments, C.P., T.J., J.K., M.F.; formal analysis, J.K., M.F.; data reduction, J.K., M.F.; writing---original draft preparation, M.F.; writing---review and editing, all authors; supervision, J.K. and M.F.; funding acquisition, J.K. and M.F. All authors have read and agreed to the published version of the manuscript.''}
\section*{Acknowledgments}
We are grateful to Dr.\,K.F. Bruder and Covestro AG for providing various \textit{Bayfol HX} films. Technical assistance at the VCN beamline PF2/VCN by Thomas Brenner is acknowledged.
This research was partially funded by the Austrian Research Promotion Agency FFG, Quantum-Austria NextPi: grant number FO999896034, European Union: NextGenerationEU. For open access purposes, the author has applied a CC BY public copyright license to any author accepted manuscript version arising from this submission. Open Access funding provided by University of Vienna.

% \abbreviations{Abbreviations}{
% The following abbreviations are used in this manuscript:\\
% 
% \noindent 
% \begin{tabular}{@{}ll}
% ILL & Institut Laue-Langevin\\
% VCN & very cold neutrons\\
% nDOE & (neutron) diffractive optical element\\ 
% \textsf{SPPS} & layers of substrate-polymer-polymer-substrate\\
% \textsf{SPSP} & layers of polymer-substrate-polymer-substrate\\
% 3CWA & 3-wave (first order) coupling analysis\\
% EMG & exponentially modified Gaussian distribution\\
% FWHM & full width at half maximum\\
% \end{tabular}
% }

\bigskip

\appendix
\section{Diffraction efficiency for the LL-interferometer (2CW, light)}
\label{sec:appendix}
We assume light diffraction and a 2CW theory based on Refs.\,\cite{Uchida-josa73,Kubota-oa78,Au-jmo87} allowing individual parameters for each layer $\ell=(1,2,3)$. To simplify notation, we map the following parameters to each other: 
\begin{itemize}
\item $n_P^\text{\scriptsize{(1)}}\to n_1$, $n_S\to n_2$, $n_P^\text{\scriptsize{(2)}}\to n_3$
\item $d_P^\text{\scriptsize{(1)}}\to d_1$, $d_S\to d_2$, $d_P^\text{\scriptsize{(2)}}\to d_3$
\end{itemize}
Then the diffraction efficiency of the first orders $\eta_{\pm 1}$ for the LL-interferometer reads:
\newcommand{\sinc}{\textrm{sinc}}
\begin{widetext}
\begin{eqnarray}
 \label{eq:2CW-3L}
 \nonumber \eta_{\pm 1}&=&\frac{c_0}{c_{\pm 1}}
\left\{\nu_3^2 \left(\cos^2\Phi_1+\xi_1^2 \sinc^2\Phi_1\right) \sinc^2\Phi_3
+\nu_1^2 \sinc^2\Phi_1 \left(\cos^2\Phi_3+\xi_3^2 \sinc^2\Phi_3\right)\right.\\
&+&2\nu_1\nu_3 \sinc\Phi_1\sinc\Phi_3 \left[\cos(2 \xi_2 +\phi )(\xi_1 \xi_3 \sinc\Phi_1 \sinc\Phi_3-\cos\Phi_1 \cos\Phi_3)\right.\\\nonumber
 & - &\left.\left.
 \sin(2 \xi_2 +\phi) (\xi_1 \cos\Phi_3 \sinc\Phi_1+\xi_3 \cos\Phi_1 \sinc\Phi_3)\right]\right\},
\end{eqnarray} 
\end{widetext}
where  
\begin{eqnarray*}
c_0&=&\cos(\theta),\\
 c_{\pm 1}&=&\sqrt{1-(\sin\theta\pm G/\beta_\ell)^2},\\
 \xi_{\ell}&=&\frac{d_{\ell}\beta_\ell(c_0-c_{\pm 1})}{2}~(\text{with}\,\ell=1,2,3),\\
 \nu_\ell&=&\frac{\beta_\ell \Delta n_{\pm 1;\ell}d_{\ell}}{2n_\ell\sqrt{c_0c_{\pm 1}}}~(\text{with}\,\ell=1,3)\,\text{, and}\\
 \Phi_\ell&=&\sqrt{\nu_\ell^2+\xi_\ell^2};\ell=1,3.
\end{eqnarray*}

Here, $G=2\pi/\Lambda$ is the spatial frequency of the gratings, $\beta=2\pi n_\ell/\lambda$ the propagation constant in the layer and $\phi$ the lateral phase shift between the two gratings (see Fig.\,\ref{fig:geo}). Angles $\theta$ are measured in the corresponding layer, accounting for their individual mean refractive indices $n_\ell$. We call $\nu$ the grating strength -- which obviously vanishes for the substrate layer -- and $\xi$ is a dephasing parameter similar to that of the ubiquitous Kogelnik notation\,\cite{Kogelnik-atj69}.

%\bibliographystyle{elsarticle-num}
\bibliography{isistr-Jab,publications_url,references24,bayfol}
\end{document}